# Colossal anomalous Stark shift in defect emission of undulated 2D materials

Sunny Gupta[1], and Boris I. Yakobson[1,2,3*]

[1]Department of Materials Science and Nanoengineering, Rice University, Houston, TX, 77005 USA
[2]Department of Chemistry, Rice University, Houston, TX 77005, USA
[3]Smalley-Curl Institute for Nanoscale Science and Technology, Rice University, Houston, TX, 77005 USA

## Abstract

We report a strikingly new physical phenomenon that mirror symmetry breaking in undulated two-dimensional (2D) materials induces a colossal Stark shift in defect emissions, occurring without external electric field ***F***, termed anomalous Stark effect. First-principles calculations of multiple defects in bent 2D *h*BN uncover the fundamental physical reasonings for this anomalous effect and reveal this arises due to strong coupling between flexoelectric polarization, and defect dipole moment. This flexo-dipole interaction, similar to that in traditional Stark effect due to ***F***, results in zero-phonon line (ZPL) shifts >500 meV for defects like $N_BV_N$ and $C_BV_N$ at $\kappa$ = 1/nm, exceeding typical Stark shifts by 2-3 orders of magnitude. The large ZPL shifts variations with curvature and bending direction offers a method to identify nanotube chirality and explain the large variability in single photon emitters' wavelength in 2D materials, with additional implications for designing nano-electro-mechanical and photonic devices.

## Introduction

Stark effect (SE) [1] – a renowned physical phenomenon causing emission lines shifts under an external electric field, has been pivotal in the early development of quantum mechanics, validating quantized electronic levels and perturbation theory, and awarded by the 1913 physics Nobel prize. This astounding phenomenon has continued to underpin new physical effects [2–13] and also found a variety of technological applications in electro-optic modulation [14–16], spin-rectification control [17], ultra-compact spectrometers [18], and tunable quantum light sources [19–22] and lasers [23]. Central to these fascinating effects is the quantum-confined SE (QCSE) which shifts photoluminescence (PL) energy by $\Delta_{PL} = -\Delta\boldsymbol{\mu}\cdot\boldsymbol{F} - \frac{1}{2}\boldsymbol{F}\cdot\Delta\boldsymbol{\alpha}\cdot\boldsymbol{F}$, governed by the electric field ($\boldsymbol{F}$), and changes in the intrinsic material parameters dipole moment ($\Delta\mu$) and polarizability ($\Delta\alpha$) between the excited and ground states. However, QCSE-induced shifts are typically limited to a few meV, posing a significant barrier for applications demanding colossal shifts, such as transitioning quantum light sources to telecom-wavelengths. On the other hand, it is fundamentally intriguing if such a renowned effect can arise in an anomalous manner, without external $\boldsymbol{F}$.

Here, we report a remarkable physical phenomenon that mirror symmetry breaking in undulated two-dimensional (2D) materials (with no Gaussian curvature, no in-plane strain) induces a colossal Stark shift in defect emissions without any external $\boldsymbol{F}$. We term this anomalous Stark effect, in a similar spirit to other anomalous effects in physics, such as anomalous Hall, anomalous quantum Hall, etc. Using first-principles calculations and examining electronic properties of multiple defects in $h$BN, a prototype 2D semiconductor, we uncover the fundamental physical reasonings for this anomalous effect and its colossal magnitude. We reveal that in undulating 2D materials, the out-of-plane flexoelectric polarization proportional to bending curvature ($\kappa$) strongly couples with the defect dipole moment. This flexo-dipole coupling $\mathscr{P} \propto \boldsymbol{\mu}\cdot\kappa$, in form similar to $\boldsymbol{\mu}\cdot\boldsymbol{F}$ in traditional SE, is colossal in strength due to the large pseudo-$\boldsymbol{F}$ arising from flexoelectricity. A large shift in zero-phonon line (ZPL) emission of >500 meV can be seen in $C_BV_N$ and $N_BV_N$ defect in bent $h$BN having $\kappa$ = 0.1/ Å, which is at least 2 - 3 orders of magnitude higher than achievable by traditional SE under an experimentally accessible external $\boldsymbol{F}$. Moreover, $\mathscr{P}$, in addition to $\kappa$ magnitude, also depends sinusoidally on the bending direction with chiral angle φ. The resulting large variation in $\Delta_{ZPL}(\kappa,\varphi)$ can be used to shift spectral lines and even identify nanotube chirality. Since undulations, like wrinkles, naturally occur in 2D materials, [24–28] this colossal $\Delta_{ZPL}(\kappa,\varphi)$ could shed light on the large variability in single photon emitters (quantum light sources) wavelength [27,29–33] in a variety of 2D materials, which otherwise were attributed to distinct defects. Our work uncovers a previously unrecognized phenomenon of anomalous and colossal Stark shift in defect emissions, and highlights their potential for designing nano-electro-mechanical sensors [34,35], highly tunable quantum light sources, and exotic photonic architectures [36].

## Results and discussion

To demonstrate the anomalous SE phenomenon, we examine various defects in 2D $h$BN, widely explored for single photon emission (SPE) [12,37,38], as a model system. $h$BN, a wide band gap semiconductor with a honeycomb lattice with two basis atoms B and N. We consider five defects $N_BV_N$, $C_BV_N$, $Li_BV_N$, $V_B$, and $V_N$, some of which have been linked to SPE [12,37,39–41], though their precise nature remains under investigation. To explore the effect of bending curvature, we use a hydrogen-passivated armchair hBN ribbon (Fig. 1a) bent along the zigzag direction to a specific $\kappa$ by fixing edge H, B, and N atoms (Fig. 1b). This setup (ribbon-model) illustrates the anomalous SE, which as shown later is a general phenomenon independent of ribbon edge or bending direction. A sufficient ribbon length of ~2.3 nm is considered (Fig. S1c), such that its band gap $E_g$ ~4.6 eV and bending stiffness $D$ = 0.84 eV agrees with that of 2D periodic hBN (Fig. S1, S2). Structural relaxation and electronic structure are calculated using density functional theory (methodology details in Supplemental Material SM-1).

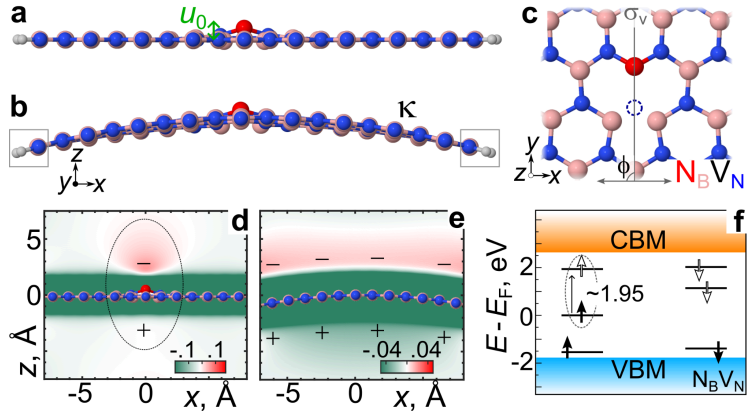

Fig. 1. (a) Relaxed structure of an armchair hBN ribbon passivated with hydrogen and containing an $N_BV_N$ defect, with the displaced nitrogen defect atom (red) protruding upward by distance $u_0$ (side view). H, B, and N atoms are shown in gray, pink, and blue, respectively. (b) Same structure as (a) in a bent geometry with curvature $\kappa$ = 0.025/Å. Edge atoms within the gray-box were fixed during structural relaxation to maintain specific curvature. (c) Top view of (a), showing the $N_BV_N$ defect with a mirror-plane $\sigma_v$ and angle $\phi$ with zig-zag (bending direction in (b)). (d) Electrostatic potential for electrons in the x-z plane (averaged over y) for the flat configuration with a defect and (e) for the bent configuration without a defect, revealing flexoelectric polarization. The color bar represents potential in volts; "+" and "−" indicating charge polarity; the ellipse in (d) highlights the point-dipole potential. (f) Electronic structure of the $N_BV_N$ defect, showing defect levels in the bandgap as black horizontal lines. The majority and minority spin states of defects are shown by up ↑ and down ↓ arrows. The states above Fermi level ($E_F$) are unoccupied and are shown by unfilled arrows. The two-level system is marked with a dashed ellipse, with an arrow marking excitation from the ground to the excited state.

We focus on the $N_BV_N$ defect to illustrate the anomalous Stark effect, though other defects also show similar behavior. $N_BV_N$ defect consists of a nitrogen substitution at a boron site adjacent to a nitrogen vacancy, with a single mirror $\sigma_v$ in the yz-plane (Fig. 1c). Its electronic structure in the ground state is shown in Fig. 1f for flat configuration, having a paramagnetic state $S$ = ½ (similar plot for bent configuration in Fig. S3c). The lowest optically allowed excitation of the defect is ~1.95 eV, marked by a dashed ellipse (Fig. 1f) and calculated optical absorption spectra (Fig. S3b), revealing a two-level system [39] capable of single-photon emission (SPE). The ZPL, which is the difference between the potential energy (PE) minimum of the defect's ground state (GS) and excited state (ES) $E_{ZPL} = E_{ES,min} - E_{GS,min}$, is routinely accessible in experiments. To calculate ZPL, we examined the ES properties by fixing the electronic levels occupancy using delta-SCF method [42] (GS level occupancy is set as zero, while ES as one) followed by structural relaxation. The calculated ZPL is ~1.42 eV (for the optical transition marked by dashed ellipse in Fig 1f), lower than the optical excitation energy (~1.95 eV) due to significant atomic relaxation in the ES.

The relaxed structure of the $N_BV_N$ defect in the GS is shown in Fig. 1a (side-view) and Fig. 1c (top-view) for the flat configuration, and in Fig. 1b (side-view) for the bent configuration. The defect in both flat and bent configuration exhibits buckling with out-of-plane displacement $u$, signifying a non-zero dipole moment $\mu_{dp}$. Interestingly, in the ES, the defect shows no buckling $u \sim 0$, in both flat and bent configurations (Fig. S4). To verify $\mu_{dp}$ in the GS, Fig. 1d shows the electrostatic potential for electrons in the flat configuration, plotted in the $x$-$z$ plane and averaged along $y$. The higher potential at the top confirms a point-like dipole with $\mu_{dp}$ pointing downward. In the bent configuration, in addition to $\mu_{dp}$, there also exists flexoelectric polarization $\mu_{fx}$. To confirm and only see $\mu_{fx}$, Fig. 1e shows the electrostatic potential for electrons (plotted similarly to Fig. 1d), for the bent ribbon without defect, revealing higher potential on the convex side, confirming $\mu_{fx}$ due to bending.

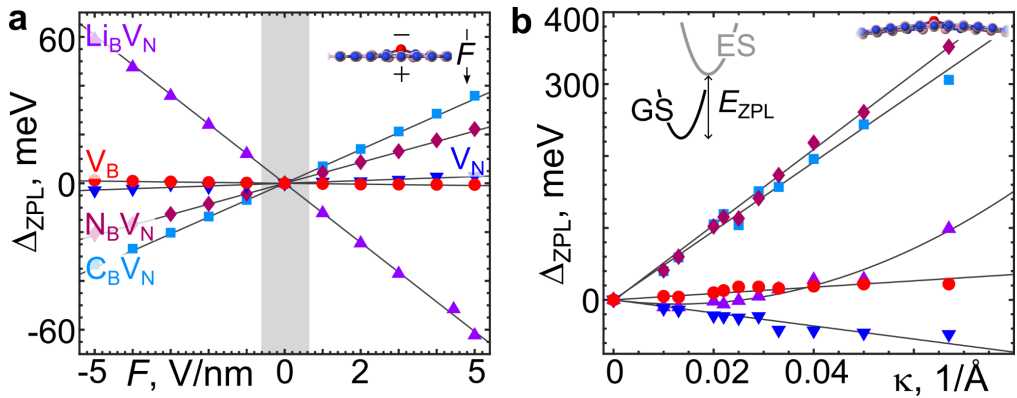

Fig. 2 (a) ZPL shifts $\Delta_{ZPL}$ for different defects in flat $h$BN ribbon under external electric field $F$ due to traditional Stark effect. Inset shows the structure of flat $h$BN with $N_BV_N$ defect and direction of external $F$. Gray region shows the experimental accessible $F$ range. (b) $\Delta_{ZPL}$ for different defects with labels same as in (a), in the bent ribbon (inset) with curvature $\kappa$ due to anomalous Stark effect. The inset also includes a schematic potential energy surface of defect's ground state (GS) and excited state (ES) and ZPL energy $E_{ZPL}$.

In the flat configuration, the difference in $\mu_{dp}$ between GS and ES, leads to a "traditional" Stark shift in ZPL emission under an external $F$, given by $\Delta_{ZPL} = -\Delta\mu_{dp} \cdot F - \frac{1}{2}F \cdot \Delta\alpha \cdot F$. In the bent configuration, however, anomalous SE due to flexo-dipole coupling is expected to cause $\Delta_{ZPL}$ solely due to curvature, without external $F$. To validate traditional SE, Fig. 2a shows calculated $\Delta_{ZPL}$ (symbols) for the lowest optically allowed transition in multiple defects under external $F$, alongside expected values (solid lines) from $\Delta_{ZPL} = -\Delta\mu_{dp} \cdot F$ (relaxed atomistic structure and electronic structure of all defects is shown in Fig. S3-S8). $\mu_{dp}$ is obtained from the charge density $\rho$ of the relaxed GS and ES structures (values shown in Table S1):

$$\mu_{dp} = \int(r - R_{center})\rho_{ions+valence}(r)d^3r \quad (1)$$

where, $r$ is the distance, $\mathbf{R}_{center}$ denotes dipole center, and $\rho$ includes both ionic and electronic charge density. The solid lines with linear slope and symbols fit well validating traditional Stark shifts (Fig. 2a). $\Delta\alpha$ is negligible for all defects. Defects like $V_B$ and $V_N$ show weak ZPL shifts due to small $\Delta\boldsymbol{\mu}_{dp}$ and lack of significant buckling (Fig. S7,S8). In contrast, $Li_BV_N$ shows the largest shift due to its high $\Delta\boldsymbol{\mu}_{dp}$, with an opposite slope to $N_BV_N$ and $C_BV_N$, reflecting the opposite sign of $\Delta\boldsymbol{\mu}_{dp}$ (Table S1) caused by Li's higher electropositivity than N and C. While large $\boldsymbol{F}$ is used in calculations, experimental $\boldsymbol{F}$ is limited to ~0.5 V/nm (gray region in Fig. 2a) [19] due to device constraints and electric breakdown, restricting traditional Stark shifts to just a few meV.

In contrast, ZPL shifts due to anomalous SE can be colossal. Fig. 2b shows $\Delta_{ZPL}$ for the same defects as Fig. 2a, caused solely by bending curvature $\kappa$, with no external $\boldsymbol{F}$. $N_BV_N$ and $C_BV_N$ defects exhibit $\Delta_{ZPL}$ >300 meV for $\kappa$ > 0.05/Å. High-curvature ($\kappa$ = 0.1/Å) hBN nanotubes have been synthesized [43], signifying even larger $\kappa$ is possible. At $\kappa$ = 0.1/Å, a $\Delta_{ZPL}$ of >0.5 eV is expected for $N_BV_N$ and $C_BV_N$. These shifts are 2-3 orders of magnitude larger than those achievable with experimentally accessible external $\boldsymbol{F}$.

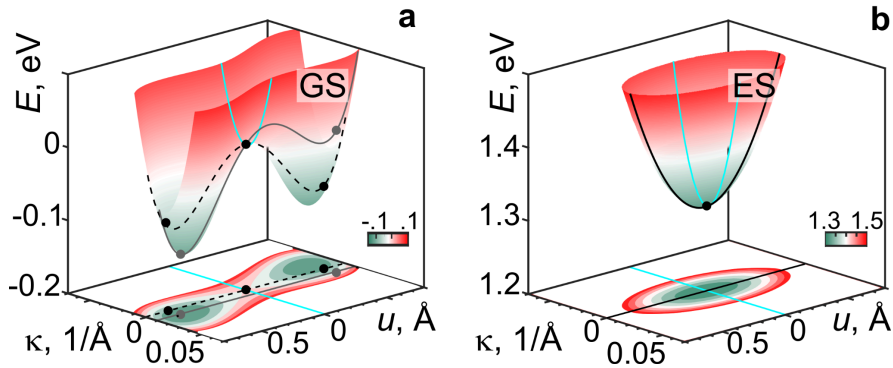

Fig. 3. Calculated potential energy surface of $N_BV_N$ defect in (a) ground (GS) and (b) excited electronic (ES) state as a function of curvature $\kappa$ and defect's out-of-plane displacement $u$. The ground state energy of the unbuckled defect in the flat layer is set as zero.

To deeply understand the physical mechanism behind the colossal ZPL shifts due to bending, we analyzed the role of flexo-dipole coupling. In the spirit of Ginzburg-Landau theories, the PE of a defect, considering out-of-plane displacement $u$, curvature $\kappa$, and relative to a unbuckled defect in flat layer, can be expressed as:

$$E(u, \kappa) = -c_1 u^2 + c_2 u^4 + \tfrac{1}{2} D A \kappa^2 + c_3 \boldsymbol{u} \cdot \kappa \quad (2)$$

where $c_{i,s}$ are constants. The first two terms form a double-well potential, typical in systems with net polarization (e.g., ferroelectrics), the third term represents bending energy of a ribbon with area $A$, and the last term captures flexo-dipole coupling between the defect dipole moment and bending curvature. The PE surface (PES) for the $N_BV_N$ defect in the GS (Fig. 3a) and ES (Fig. 3b) was fitted to this Eq. 2 using DFT-calculated energies. The GS relaxed structure in flat configuration has $u$, $\boldsymbol{\mu}_{dp} \neq 0$ (Fig. 1a), hence the PES cut along $\kappa = 0$ line (Fig. 3a, black dashed-line) depicts a double-well potential. However, a cut along finite $\kappa$ (Fig. 3a, gray solid-line) depicts asymmetry in the potential well, arising due to flexo-dipole coupling. The PES minima at $\kappa \neq 0$ suggests the defect can induce bending in a 2D material, consistent with experimental observations [44], but neglected heretofore in first-principles based defect calculations. In the ES ($u$, $\mu_{dp} \sim 0$), the PES (Fig. 3b) simplifies to $E_{ES} = c_1 u^2 + \frac{1}{2}DA\kappa^2 + 1.32$, shifted by 1.32 eV relative to GS, with no double-well potential or flexo-dipole coupling. The PES calculation confirms the presence of flexo-dipole coupling in the GS of the $N_BV_N$ defect.

The flexo-dipole coupling in Eq. 2, $\mathscr{P} \propto \boldsymbol{u} \cdot \kappa$ or $\boldsymbol{\mu}_{dp} \cdot \kappa$ ($\boldsymbol{u}$ and $\boldsymbol{\mu}_{dp}$ are equivalent), resembles the form $\boldsymbol{\mu}_{dp} \cdot \boldsymbol{F}$, which describes the PE contribution of a dipole in an external electric field. In traditional SE, this coupling results in a ZPL shift $\Delta_{ZPL}(\boldsymbol{F}) \propto -\Delta\boldsymbol{\mu}_{dp} \cdot \boldsymbol{F}$. Similarly, flexo-dipole coupling causes a ZPL shift without any external $\boldsymbol{F}$: $\Delta_{ZPL}(\kappa) = E_{ZPL}(\kappa \neq 0) - E_{ZPL}(\kappa = 0) \propto u_0 \kappa + O(\kappa^2)$, $u_0$ is the atom displacement in the relaxed flat configuration (Fig. 1a) (more details in SM-2). The solid lines in Fig. 2b show a polynomial fit $\Delta_{ZPL}(\kappa) \propto \kappa + \kappa^2$ to DFT data (symbols), showing agreement. Due to the similar form and effect of flexo-dipole coupling and dipole–electric field coupling, we term the ZPL shifts caused by bending curvature as an anomalous Stark effect, analogous to other anomalous effects in physics that arise without external fields.

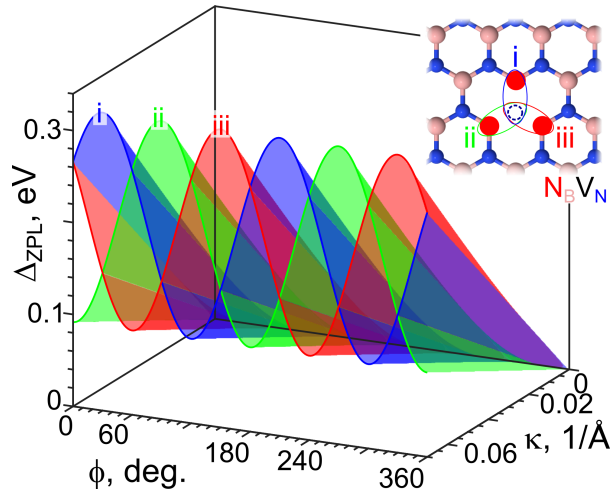

Fig. 4. Calculated $\Delta_{ZPL}$ of $N_BV_N$ defect as a function of curvature $\kappa$ and bending direction with angle $\phi$ relative to the defect's mirror line. i, ii, and iii denote three configurations of $N_BV_N$ defect (inset)

Unlike traditional SE with small ZPL shifts under experimentally accessible external $\boldsymbol{F}$ (Fig. 2a), the colossal $\Delta_{ZPL}(\kappa)$ due to anomalous SE (Fig. 2b) arises from strong flexo-dipole coupling. To ascertain the magnitude of such coupling we looked into the associated pseudo-electric field ($F_{flexo}$) due to flexoelectricity. Bending as shown in Fig. 1e creates flexoelectric polarization, which can be viewed phenomenologically as the polarization response to $F_{flexo}$, a pseudo-field [29–31]. Such pseudo-field description due to flexoelectricity has successfully explained the skewing of the double-well potential in ferroelectrics under strain gradient [45,47], mechanical polarization switching [48], and polarization control in a inhomogeneously strained metal [46]. To quantify $F_{flexo}$, we calculated $\Delta_{ZPL}$ under external $\boldsymbol{F}$ for structures with different fixed $\kappa$. $F_{flexo}(\kappa)$ was determined as the external field required to counter the effect due to bending curvature, that is, $\boldsymbol{F}$ required to make $\Delta_{ZPL}(\kappa)$ zero (Fig. S9a), for each fixed $\kappa$. For the $N_BV_N$ defect, $F_{flexo} \sim 8$ V/Å at $\kappa = 0.05$/Å (Fig. S9b), far exceeding the breakdown voltage of $h$BN and three to four orders of magnitude larger than experimentally accessible $\boldsymbol{F}$.

This extraordinary $F_{flexo}$ drives the strong flexo-dipole coupling, causing the colossal anomalous Stark shifts.

In flat $h$BN, the $N_BV_N$ defect can exist in three degenerate configurations (inset Fig. 4), with identical ZPL shifts under external $F$ (Fig. S10). However, bending lifts this degeneracy, and $\Delta_{ZPL}$ depends on the bending direction with angle $\phi$ relative to the mirror line (Fig. 1c). We computed $\Delta_{ZPL}(\kappa,\phi)$ considering various ribbon geometries and bending directions (Fig. S11). The DFT-computed data fits well to $\Delta_{ZPL}(\kappa,\phi) = c_1 \cdot \kappa \cdot \sin(2\phi-\pi/2+\phi_0) + c_2 \cdot \kappa$ (Fig. S11), where $c_i$'s are constants, and $\phi_0 = 0°$, $120°$, and $240°$ correspond to the defect configurations i, ii, and iii (inset Fig. 4), respectively, matching their relative mirror line angles. The fitted $\Delta_{ZPL}(\kappa,\phi)$ (Fig. 4) shows large ZPL variability with a periodicity of $180°$. While $F_{flexo}$ depends only on $|\kappa|$, the sinusoidal dependence of $\Delta_{ZPL}$ on $\phi$ reflects the angular dependence of flexo-dipole coupling. Such sinusoidal dependence is also observed in the formation energy of Stone-Wales defects in carbon nanotubes under strain, linked to their mechanical strength. [49,50] Since bending direction is also the direction determining the chirality in a nanotube geometry, the $N_BV_N$ defect's $\Delta_{ZPL}(\kappa,\phi)$ variation offers a method to identify nanotube chirality, typically a challenging task.

Bending curvature naturally occurs in 2D materials, such as wrinkles [24,26–28,51], due to 2D material's large Föppl–von Kármán number [26,52]. Our predicted colossal Stark shift suggests that defects on such curved surfaces will show significant ZPL emission variability based on local curvature. Experiments have shown substantial emission wavelength variability from SPE sources in $h$BN [27,29–33], many of which were reported to also have wrinkles [27,30,32]. This variability has been speculatively attributed to different defects, however, it could also stem from a single defect type in different curved regions. Nonetheless, the role of curvature and the induced anomalous Stark shift in ZPL emission has been overlooked until now. The anomalous Stark shift must be accounted for in experiments to accurately identify SPE sources in $h$BN, a critical step and significant challenge for their practical application.

In summary, we report a heretofore unrecognized phenomenon that undulations in 2D $h$BN induce a colossal anomalous Stark shift in defect emissions, driven by flexo-dipole coupling. This effect extends to other 2D material defect combinations [53] and can be evaluated in a high-throughput manner using our ribbon-model. Flexo-dipole coupling also enables defects to induce local bending in 2D materials, a critical factor often overlooked in first-principles-based defect calculations but essential for accurately comparing theoretical results with experiments. In certain systems with designed curvature, the colossal ZPL shift due to anomalous SE, can tune SPE to even optical telecommunications bands, which is of utmost practical importance and is challenging using external electric fields. Additionally, the large ZPL shifts due to curvature can be used to design nano-opto-electro-mechanical [34,35] devices and sensors, with enhanced sensitivity and resolution.